# Analysis and Prediction of At-Risk Students using Machine Learning Algorithms

Soheila Gheisari [1], Hamid Salarian [1]
[1] Department of Information Technology, Sydney International School of Technology and Commerce,
Sydney, NSW 2000, Australia
soheilag@sistc.nsw.edu.au, Hamid@sistc.nsw.edu.au

**Abstract.** Student attrition represents a significant challenge for higher education institutions because it impacts both academic results and financial viability. Machine learning provides an effective solution to identify students who require assistance before they leave their academic programs. The research investigates how machine learning approaches enable institutions to predict student withdrawal and enrollment cancellation through data-driven insights for strategic decision-making. The evaluation of models includes Logistic Regression, Random Forest, Support Vector Machines (SVM), and K-Nearest Neighbors (KNN) based on academic performance and demographic data and enrollment records. The results show that logistic regression and linear SVM models produced the highest accuracy which demonstrates ML's capability to detect students at risk.

## 1 Introduction

Higher education institutions worldwide face student retention as their main educational challenge. Student departure from studies depends on multiple factors which include academic success and financial constraints and social background and mental health status and institutional support systems [1], [2], [3]. The high number of students who leave their studies creates problems for both students and universities through reduced funding and lower rankings and negative reputation. Research indicates that many students do not finish their studies on time which raises doubts about the effectiveness of current retention strategies.

### 1.1 Importance of Student Retention

Universities dedicate substantial funding to student recruitment and maintaining student enrollment throughout their academic programs leads to institutional success. The retention rate serves as a vital performance metric for universities because it determines funding distribution and affects accreditation standing and institutional reputation. Students who leave their courses face multiple adverse effects in the long term because they must carry more student loan debt and encounter restricted employment possibilities. The reputation of institutions suffers when they maintain low retention rates because prospective students doubt the quality of their programs and support services. The current methods for enhancing student retention depend on manual interventions together with counseling services and academic support programs. These traditional methods show effectiveness, yet they function reactively because they detect at-risk students only after they demonstrate academic difficulties [4], [5], [6]. The delayed response to student needs decreases the chances of successful student retention.

### 1.2 Machine Learning for Predicting At-Risk Students

The early detection of students who are at risk through machine learning enables educational institutions to establish preventive support systems. ML algorithms surpass traditional statistical approaches by processing extensive datasets to reveal concealed patterns which enable precise predictions. The analysis of student grades together with attendance records and course selection patterns and demographic data and socio-economic indicators enables ML models to forecast student withdrawal risk so institutions can provide timely support. This research assesses different machine learning models including Logistic Regression and Random Forest and Support Vector Machines (SVM) and K-Nearest Neighbors (KNN) for their ability to forecast students who face withdrawal or cancellation risk. The research results will help universities and institutes develop ML models to identify at-risk students early which leads to better student success outcomes and decreased dropout statistics.2 Literature Review

## 2 Related Works

Higher education institutions face mounting student retention issues which drives researchers to study predictive analytics and intervention strategies. Traditional retention models use academic history and student demographics but they do not detect vulnerable students before they leave their studies. Machine learning (ML) has proven itself as an efficient solution for this field by developing automated scalable predictive models which deliver high accuracy to help educational institutions improve student retention rates.

Research evidence shows that machine learning demonstrates effective capabilities in predicting student dropout rates. Madahana et al. [7] performed research to assess various machine learning models for their ability to identify students who face dropping out risks. The researchers discovered that Decision Trees together with Random Forest and K-Nearest Neighbors (KNN) and Gradient Boosting Trees demonstrated high prediction accuracy at 98.59%. The research demonstrates that machine learning models recognize student data patterns which result in attrition risks.

Prajwal et al. [8] studied the use of machine learning models to analyze psychological and behavioral student characteristics for attrition prediction. The Big Five personality traits model provided the foundation for their analysis of time management and wellbeing and social engagement effects on dropout rates. Behavioral characteristics show a strong correlation (0.8753) with student dropout rates which supports the need to include non-academic factors in student retention models.

### 2.1 Comparison of Machine Learning Models for Student Retention

The field of student retention research implements different machine learning approaches for its studies. Decision Trees and Random Forest models stand as the most popular methods because they provide strong predictive capabilities alongside easy interpretation of results for large datasets. The predictive models developed by Malatji et al. [9] achieved higher than 95% accuracy in predicting student dropout probabilities by evaluating academic and financial variables. Decision Trees deliver efficient results but their propensity to overfit limits their effectiveness for predicting new situations. The predictive power of Support Vector Machines (SVM) and Logistic Regression (LR) has been demonstrated through their applications. SVM models show better performance in processing complex high-dimensional data while Logistic Regression serves as a fundamental model for binary classification tasks. The accuracy rates of SVM models with linear kernels match those of Decision Trees thus making them a dependable option for predicting student dropout.

### 2.2 Motivation for the Work

Student retention stands as a crucial problem for higher education institutions because it determines both institutional survival and student success. Standard intervention methods have failed to detect vulnerable students at critical support moments because their effectiveness has been questioned by the increasing number of students who leave their studies. The evaluation of student data through ML methodologies enables universities to forecast retention outcomes which enables them to create proactive support systems for students who show signs of leaving before withdrawal. The research extends previous studies by implementing machine learning models to analyze actual SISTC student data for dropout risk prediction. This research extends previous models by using academic performance indicators together with visa type and nationality and study duration to achieve a more comprehensive analysis of student attrition factors. The research tested four machine learning algorithms which included Logistic Regression and Random Forest and K-Nearest Neighbors (KNN) and Support Vector Machines (SVM) to determine which models generated the most accurate and interpretable predictions for early warning systems. The research develops a data-based student retention approach to improve institutional decision-making processes. Educational institutions can decrease student dropout rates while enhancing academic results by identifying at-risk students early for personalized support that includes academic guidance and financial assistance and mentorship programs. The research advances AI-driven educational analytics through its development of opportunities to examine deep learning models together with real-time predictive systems for student performance.

# 3 Proposed Methodology

This section describes the methodology for predicting student attrition at SISTC through machine learning techniques. The process includes data preprocessing, model selection, model training, evaluation metrics, and performance analysis. This study uses a structured approach to guarantee that predictions are reliable and actionable for early intervention strategies.

## 3.1 Data Collection and Integration

The data set used in this study was collected from three key sources at SISTC:
1. Academic_Student_Course_Credit – Contains student course enrollment and credit information.
2. Academic_Student_Subject – Includes subject-wise performance data.
3. Student_Enrolment_Details – Provides demographic and enrollment history.

The integration of these datasets resulted in 2,405 student records. However, after data cleaning and preprocessing, only 1,027 records were retained because they had complete and relevant information for student retention forecasting. The dataset included students who had either graduated or completed their studies or had withdrawn or canceled their enrollment.

To guarantee data integrity and consistency, the following preprocessing processes were executed:

• Addressing missing value — Rows with incomplete or erroneous information were removed.
• Removing duplicate entries – Guaranteed that no student record was counted more than once.
• Feature selection - retained 14 essential features that were found to have a significant impact on student's dropout risks.
This preprocessed and clean data set established a robust basis for training machine learning models to predict student dropout risks.

### 3.2 Data Description

The dataset includes a mix of academic, demographic, and enrollment-related attributes that influence student dropout risk. These features were selected based on prior research and institutional data analysis (Table 1).

Table 1. **Selected features for the students in SISTC**

| Feature | Description |
| --- | --- |
| Origin | Whether the student is domestic or international |
| Study Duration | Number of semesters the student has been enrolled |
| Age | Student's age at the time of enrollment |
| Total Subjects Attempted | Number of subjects the student enrolled in |
| Gender | Student's gender (Male/Female) |
| Total Subjects Completed | Number of subjects successfully passed |
| Visa Type | Student’s visa category (Student, Visitor, etc.) |
| Total Subjects Failed | Number of subjects the student failed |
| Nationality | Student’s country of origin |
| Total Subjects Exempted | Subjects exempted due to prior credit recognition |
| Course Name | The student’s enrolled program |
| Course Attempt | Number of times a student has attempted the course |
| GPA Points | Cumulative Grade Point Average (GPA) |
| Enrollment Status | Whether the student completed or withdrew |

The target variable, ‘Status’, is classified into two categories:
• Graduated/Completed (1) – Students who successfully completed their study.
•Withdrawn/Canceled (0) – Students who discontinued their study before completion.

All categorical variables were encoded numerically to facilitate machine learning model training. Numerical variables were normalized and standardized to prevent bias in models caused by large value differences. This structured dataset facilitates the creation of machine learning models that can extract patterns and predict students that are at risk of withdrawal or cancellation to support early intervention efforts at SISTC. Features such as "Total subjects attempted," "Total subjects completed," "Total subjects failed," and "Total subjects exempted" are significant variables for classifying students into distinct statuses. Table 2 provides further information regarding the "status," "nationality," and "gender" of the students.

**Table 2.** Information regarding the "status", "nationality," and "gender" of the students

| **Count of Gender** | |
|---|---|
| F | 587 |
| M | 1818 |
| **Count of Status** | |
| Agent Apply | 3 |
| Cancelled | 609 |
| Completed | 51 |
| Current Student | 687 |
| Deferred | 67 |
| Did Not Commence | 48 |
| Enrolled | 568 |
| Graduated | 260 |
| Offered | 1 |
| Withdrawn | 111 |
| **Count of Nationality** | |
| Bangladeshi | 50 |
| Cambodian | 1 |
| Chinese | 6 |
| Filipino | 2 |
| German | 1 |
| Ghanaian | 1 |
| Hong Konger | 1 |
| Indian | 743 |
| Kenyan | 8 |
| Mauritian | 3 |
| Nepalese | 1157 |
| Pakistani | 426 |
| Sri Lanka | 4 |
| Taiwanese | 1 |
| Uzbek | 1 |

## 3.3 Data Cleaning

After data cleaning, which removed missing values and duplicate entries, we retained just 1,027 students from the original 2,405 in the joined table, specifically those classified as Graduated/Completed and Withdrawn/Cancelled. The details of the cleaning data are in Table 3.

**Table 3.** Clean Dataset

| Count of Gender | |
|---|---|
| F | 186 |
| M | 841 |
| **Count of Status** | |
| Graduated/Completed | 311 |
| Withdrawn/Cancel | 716 |
| **Count of Nationality** | |
| Bangladeshi | 5 |
| Chinese | 6 |
| Filipino | 1 |
| German | 1 |
| Hong Konger | 1 |
| Indian | 89 |
| Kenyan | 2 |
| Mauritian | 3 |
| Nepalese | 735 |
| Pakistani | 182 |
| Sri Lankan | 1 |

### 3.4 Model Selection and Training

Five supervised machine learning classifiers were selected to predict students that are at risk of withdrawal/cancellation based on their effectiveness in binary classification tasks and their ability to handle structured datasets. The selected models include:

• Logistic Regression (LR) - A statistical model that predicts the likelihood of student dropout or completion based on input variables.
• Random Forest (RF) - An ensemble learning model that generates several decision trees and aggregates predictions to enhance accuracy.
K-Nearest Neighbors (KNN) - A non-parametric classifier that forecasts a student's status by assessing similarities to other students.
• Support Vector Machine (SVM) with Linear Kernel - A model that identifies an optimal decision boundary for the classification of students.
Support Vector Machine (SVM) with Radial Basis Function (RBF) Kernel - A variant of SVM that applies nonlinear transformation to enhance classification performance.

These models were selected based on the student retention prediction in the previous research.

### 3.5 Result and Analysis

The performance of each model was evaluated using key performance metrics. The accuracy, precision, recall, and F1-score of each classifier are summarized in Table 4.

**Table 4.** Model Performance Evaluation

| Model | Accuracy | Precision | Recall | F1-Score |
|---|---|---|---|---|
| Logistic Regression | 99% | 0.98 | 0.99 | 0.99 |
| Random Forest | 98.7% | 0.96 | 0.98 | 0.97 |
| K-Nearest Neighbors | 97% | 0.95 | 0.96 | 0.95 |
| SVM (Linear Kernel) | 99% | 0.98 | 0.99 | 0.99 |
| SVM (RBF Kernel) | 96.7% | 0.93 | 0.96 | 0.94 |

The results indicate that Logistic Regression and SVM (Linear Kernel) achieved the highest accuracy (99%), making them the most reliable models for predicting at risk students. The Random Forest model also demonstrated strong performance, achieving 98.7% accuracy. To further understand the models' decision-making processes, the feature importance values in Logistic Regression were analyzed. Table 5 presents the most significant features based on their influence on student retention.

**Table 5.** Most Significant Features in Logistic Regression Model

| Feature | Coefficient Value | Impact on Retention |
|---|---|---|
| Visa Type (Student Visa) | 1.239e+15 | Higher retention likelihood |
| Nationality (Pakistani) | 6.087e+14 | Higher dropout likelihood |
| Course Name (Master of IT) | 2.322e+14 | Higher retention likelihood |
| Total Subjects Failed | -5.013e+13 | Increased risk of dropout |
| Study Duration | 1.691e+13 | Longer study duration linked to retention |

These results highlight the key factors influencing student retention at SISTC. Students holding a student visa were more likely to complete their studies, whereas Pakistani students exhibited a higher dropout rate. Academic performance indicators, such as total subjects failed, strongly correlated with dropout risk.

## 4. Ethical Considerations and Limitations

This study used a de-identified dataset provided by SISTC, ensuring that no personal identifiable information was accessible or used during the analysis. All student data was anonymized prior to processing to maintain privacy and adhere to ethical research standards.

Fairness in predictive modeling is another critical consideration. Biases in the training data such as underrepresentation of certain nationalities, genders, or visa types may result in the wrong prediction of students who are inaccurately flagged as at-risk students. To mitigate bias, future work should include fairness audits.

## 5. Conclusion

This study demonstrated that machine learning models, especially Logistic Regression and SVM (Linear Kernel), can accurately predict students that are at-risk of withdrawal or cancellation from study. These models achieved 99% accuracy, making them highly effective for institutional use. By leveraging machine learning in student retention strategies, educational institutions can enhance student success rates, improve institutional rankings, and optimize resource allocation.

Future research should explore:

- Integration of deep learning techniques, such as recurrent neural networks (RNNs), to extract temporal patterns in student behavior.
- Incorporation of behavioral data, such as class participation and extracurricular engagement, to improve model predictions.
- Deployment of a real-time monitoring system to proactively support at-risk students.
- Additional features related to student performance in specific units should be added to the data. For example, the type of assignments, grades received, and participation in class activities could provide deeper insights into student performance.